# KMT-2022-BLG-0440Lb: A New $q < 10^{-4}$ Microlensing Planet with the Central-Resonant Caustic Degeneracy Broken

Jiyuan Zhang[1], Weicheng Zang[1,2]★, Youn Kil Jung[3,4], Hongjing Yang[1]†, Andrew Gould[5,6], Takahiro Sumi[7], Shude Mao[1,8], Subo Dong[9]
(Leading Authors)
Michael D. Albrow[10], Sun-Ju Chung[3,4], Cheongho Han[11], Kyu-Ha Hwang[3], Yoon-Hyun Ryu[3], In-Gu Shin[2], Yossi Shvartzvald[12], Jennifer C. Yee[2], Sang-Mok Cha[3,13], Dong-Jin Kim[3], Hyoun-Woo Kim[3], Seung-Lee Kim[3], Chung-Uk Lee[3], Dong-Joo Lee[3], Yongseok Lee[3,13], Byeong-Gon Park[3,4], Richard W. Pogge[6]
(The KMTNet Collaboration)
Qiyue Qian[1], Zhuokai Liu[9], Dan Maoz[14], Matthew T. Penny[15], Wei Zhu[1]
(The MAP & μFUN Follow-up Teams)
Fumio Abe[16], Richard Barry[17], David P. Bennett[17,18], Aparna Bhattacharya[17,18], Ian A. Bond[19], Hirosane Fujii[16], Akihiko Fukui[20,21], Ryusei Hamada[7], Yuki Hirao[7], Stela Ishitani Silva[22,17], Yoshitaka Itow[16], Rintaro Kirikawa[7], Iona Kondo[7], Naoki Koshimoto[17,18], Yutaka Matsubara[16], Sho Matsumoto[7], Shota Miyazaki[7], Yasushi Muraki[16], Arisa Okamura[7], Greg Olmschenk[17], Clément Ranc[23], Nicholas J. Rattenbury[24], Yuki Satoh[7], Daisuke Suzuki[7], Taiga Toda[7], Mio Tomoyoshi[7], Paul J. Tristram[25], Aikaterini Vandorou[17,18], Hibiki Yama[7], Kansuke Yamashita[7]
(The MOA Collaboration)

[1] *Department of Astronomy, Tsinghua University, Beijing 100084, China*
[2] *Center for Astrophysics | Harvard & Smithsonian 60 Garden St., Cambridge, MA 02138, USA*
[3] *Korea Astronomy and Space Science Institute, Daejon 34055, Republic of Korea*
[4] *University of Science and Technology, Korea, (UST), 217 Gajeong-ro Yuseong-gu, Daejeon 34113, Republic of Korea*
[5] *Max-Planck-Institute for Astronomy, Königstuhl 17, 69117 Heidelberg, Germany*
[6] *Department of Astronomy, Ohio State University, 140 W. 18th Ave., Columbus, OH 43210, USA*
[7] *Department of Earth and Space Science, Graduate School of Science, Osaka University, Toyonaka, Osaka 560-0043, Japan*
[8] *National Astronomical Observatories, Chinese Academy of Sciences, Beijing 100101, China*
[9] *Kavli Institute for Astronomy and Astrophysics, Peking University, Yi He Yuan Road 5, Hai Dian District, Beijing 100871, China*
[10] *University of Canterbury, Department of Physics and Astronomy, Private Bag 4800, Christchurch 8020, New Zealand*
[11] *Department of Physics, Chungbuk National University, Cheongju 28644, Republic of Korea*
[12] *Department of Particle Physics and Astrophysics, Weizmann Institute of Science, Rehovot 76100, Israel*
[13] *School of Space Research, Kyung Hee University, Yongin, Kyeonggi 17104, Republic of Korea*
[14] *School of Physics and Astronomy, Tel-Aviv University, Tel-Aviv 6997801, Israel*
[15] *Department of Physics and Astronomy, Louisiana State University, Baton Rouge, LA 70803 USA*
[16] *Institute for Space-Earth Environmental Research, Nagoya University, Nagoya 464-8601, Japan*
[17] *Code 667, NASA Goddard Space Flight Center, Greenbelt, MD 20771, USA*
[18] *Department of Astronomy, University of Maryland, College Park, MD 20742, USA*
[19] *Institute of Natural and Mathematical Sciences, Massey University, Auckland 0745, New Zealand*
[20] *Department of Earth and Planetary Science, Graduate School of Science, The University of Tokyo, 7-3-1 Hongo, Bunkyo-ku, Tokyo 113-0033, Japan*
[21] *Instituto de Astrofísica de Canarias, Vía Láctea s/n, E-38205 La Laguna, Tenerife, Spain*
[22] *Department of Physics, The Catholic University of America, Washington, DC 20064, USA*
[23] *Sorbonne Université, CNRS, Institut d'Astrophysique de Paris, IAP, F-75014, Paris, France*
[24] *Department of Physics, University of Auckland, Private Bag 92019, Auckland, New Zealand*
[25] *University of Canterbury Mt. John Observatory, P.O. Box 56, Lake Tekapo 8770, New Zealand*

2 May 2023

**ABSTRACT**

We present the observations and analysis of a high-magnification microlensing planetary event, KMT-2022-BLG-0440, for which the weak and short-lived planetary signal was covered by both the KMTNet survey and follow-up observations. The binary-lens models with a central caustic provide the best fits, with a planet/host mass ratio, $q = 0.75$–$1.00 \times 10^{-4}$ at $1\sigma$. The binary-lens models with a resonant caustic and a brown-dwarf mass ratio are both excluded by $\Delta\chi^2 > 70$. The binary-source model can fit the anomaly well but is rejected by the "color argument" on the second source. From Bayesian analyses, it is estimated that the host star is likely a K or M dwarf located in the Galactic disk, the planet probably has a Neptune-mass, and the projected planet-host separation is $1.9^{+0.6}_{-0.7}$ or $4.6^{+1.4}_{-1.7}$ au, subject to the close/wide degeneracy. This is the third $q < 10^{-4}$ planet from a high-magnification planetary signal ($A \gtrsim 65$). Together with another such planet, KMT-2021-BLG-0171Lb, the ongoing follow-up program for the KMTNet high-magnification events has demonstrated its ability in detecting high-magnification planetary signals for $q < 10^{-4}$ planets, which are challenging for the current microlensing surveys.

# 1 INTRODUCTION

Because the source trajectory of a high-magnification microlensing event goes close to the host star, where every planet induces distortions in the magnification profile by their central or resonant caustics (Griest & Safizadeh 1998), high-magnification events are sensitive to planets and play an important role in microlensing planet detections. For example, among the five unambiguous multi-planetary systems detected by microlensing (Gaudi et al. 2008; Han et al. 2013, 2019, 2022a,b), all were detected by central or resonant caustics and four had $A_{\rm max} > 80$ for the underlying single-lens events. High-magnification events are good targets for follow-up observations because the peaks are predictable and are often bright enough for small telescopes. Follow-up observations for high-magnification events can form a statistical sample. Using the 13 homogeneously-selected high-magnification events ($A_{\rm max} > 200$) observed by the Microlensing Follow Up Network ($\mu$FUN), Gould et al. (2010) formed a statistical sample of six planets and presented the first measurement of the planet frequency beyond the snow line.

The current largest published statistical sample of microlensing planets (Suzuki et al. 2016) has 22 planets, with only two $q < 10^{-4}$ planets ($q$ is the planet/host mass ratio) and no two-planetary systems. To build a larger statistical sample, since July 2020 the Microlensing Astronomy Probe (MAP[1]) collaboration has been conducting a long-term follow-up program for high-magnification events using the Las Cumbres Observatory (LCO) global network (Brown et al. 2013). The program also contains follow-up observations from $\mu$FUN and the "auto-followup" system of the Korea Microlensing Telescope Network (KMTNet, Kim et al. 2016). Despite the considerable difficulties imposed by Covid-19 in 2020, this program detected the lowest-mass-ratio ($q = 0.9$–$1.2 \times 10^{-5}$ at $1\sigma$) microlensing planet to date in the event KMT-2020-BLG-0414, with a second companion at the planet/brown-dwarf boundary (Zang et al. 2021a).

In the 2021 season, this follow-up program detected at least six planets. Yang et al. (2022) found that two of them, KMT-2021-BLG-0171Lb and KMT-2021-BLG-1689Lb, suffer from the "central-resonant" caustic degeneracy, for which the short-lived bump-type planetary signals can be respectively fitted by a central-caustic model and a resonant-caustic model. As shown in Figures 3 and 4 of Yang et al. (2022), the differences between the two models are short ($< 0.1$ days) and weak ($\Delta I < 0.05$ mag). Although the differences were covered by dense $\mu$FUN data, the degeneracy cannot be fully resolved due to insufficient photometric accuracy. In the same season, the high-cadence ($\Gamma \geqslant 2$ hr$^{-1}$) KMTNet data have found another three planetary events which have the "central-resonant" caustic degeneracy (Ryu et al. 2022; Shin et al. 2023), but none of them have the degeneracy been broken with a significance level of $> 5\sigma$. Therefore, the "central-resonant" caustic degeneracy should be common and it probably requires high-cadence high-accuracy follow-up data to break it.

Here we report the analysis of a high-magnification planetary event from the 2022 season, KMT-2022-BLG-0440. The planetary signal is also a short-lived bump, but the KMTNet data and the LCO follow-up data break the "central-resonant" caustic degeneracy. The paper is structured as follows. In Section 2, we introduce the survey and follow-up data of this event. In Section 3, we conduct the binary-lens single-source (2L1S) analysis. In Section 4 we show the color-magnitude diagram analysis. Because a short-lived bump can also be caused by a single-lens binary source (1L2S) model, we present the 1L2S analysis in Section 5. In Section 6 we estimate the physical parameters of the planetary system. Finally, we investigate the results only using the survey data and discuss the implications of this work in Section 7.

# 2 OBSERVATIONS AND DATA REDUCTION

## 2.1 Survey Observations

Figure 1 displays all of the light curves acquired for the microlensing event, KMT-2022-BLG-0440, which was first discovered by the KMTNet AlertFinder system on 15 April 2022 (Kim et al. 2018a). The KMTNet data were taken from three identical 1.6 m telescopes equipped with 4 deg$^2$ cameras (Kim et al. 2016) at the Cerro Tololo Inter-American Observatory (CTIO) in Chile (KMTC), the South African Astronomical Observatory (SAAO) in South Africa (KMTS), and the Siding Spring Observatory (SSO) in Australia (KMTA). The event is located at equatorial coordinates of $(\alpha, \delta)_{\rm J2000}$ = (17:58:20.06, −32:17:43.12), corresponding to Galactic coordinates of $(\ell, b) = (-1.50, -4.06)$. It lies in the KMTNet BLG22 and BLG41 fields. See Figure 12 of Kim et al. (2018b) for the field placement. The nominal cadences of the (BLG22, BLG41) field are $(1.0, 2.0)$ hr$^{-1}$ for KMTC, and $(0.75, 1.5)$ hr$^{-1}$ for KMTS and KMTA. Most of the KMTNet images were taken in the $I$ band for the light-curve analysis, and a fraction of $V$-band images were acquired for source color measurements. For the current case, the $V$-band data of the BLG41 field taken from KMTC and KMTS help exclude the 1L2S model, so we include them in the light-curve analysis.

⋆ E-mail: 3130102785@zju.edu.cn
† E-mail: hongjing.yang@qq.com
[1] http://i.astro.tsinghua.edu.cn/~smao/MAP/

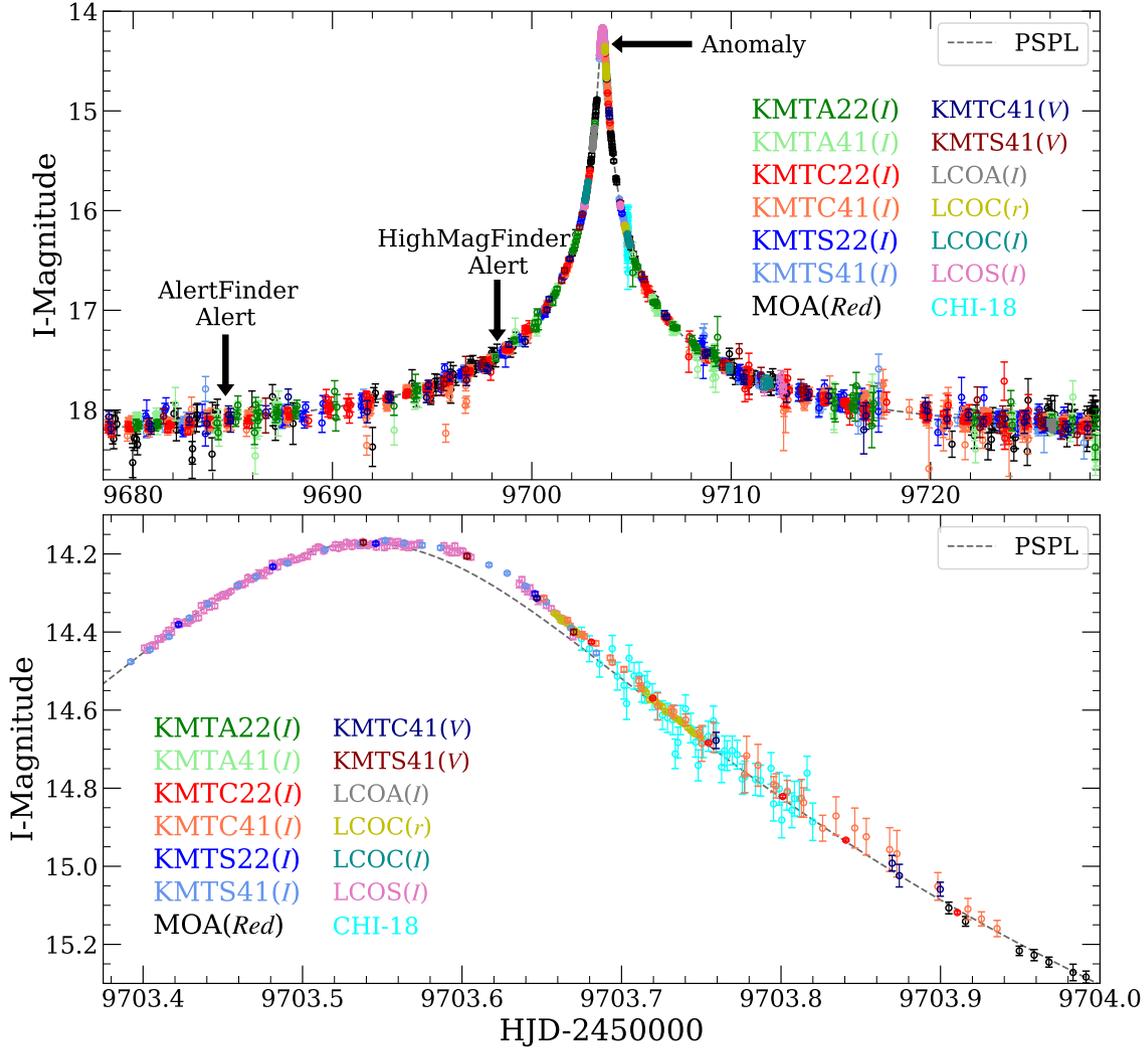

**Figure 1.** Light curve of the microlensing event, KMT-2022-BLG-0440. The colors of the data points are set to match those of the data sets marked in the legend. The dashed line shows the best-fit point-source point-lens (PSPL) model. The three arrows indicate the moments when the event was found by the KMTNet AlertFinder system, was alerted by the KMTNet HighMagFinder system, and showed an anomaly, respectively. The $I$-band KMTC41 images were affected by a bleed trail from a saturated star and the resulting photometry is inconsistent with other data sets, so the adopted results were obtained by excluding the KMTC41 ($I$) data.

The event was later identified by the Microlensing Observations in Astrophysics (MOA, Sako et al. 2008) group as MOA-2022-BLG-199 on 27 April 2022 (Bond et al. 2001). Hereafter, we designate the event by its first-discovery name, KMT-2022-BLG-0440. The MOA group conducted observations using one 1.8 m telescope equipped with a 2.2 deg$^2$ camera at the Mt. John University Observatory in New Zealand. The nominal cadence for the MOA observations is $\sim 0.7$ hr$^{-1}$ on average. The MOA images were mainly acquired in the MOA-Red band, which is similar to the sum of the standard Cousins $R$- and $I$-band filters.

The event is also located at the BLG507 and BLG508 fields of the fourth phase of the Optical Gravitational Lensing Experiment (OGLE, Udalski et al. 2015). However, due to the Covid-19 pandemic, OGLE stopped regular observations in March 2020 and resumed in August 2022. Thus, OGLE only observed the baseline of this event and we do not include the OGLE data in the light-curve analysis.

### 2.2 Follow-up Observations

At UT 18:18 on 28 April 2022 (HJD′ = 9698.263, HJD′ = HJD − 2450000), the KMTNet HighMagFinder system (Yang et al. 2022) found that this event could peak at a high magnification about five days later. Following the alert, three groups conducted follow-up observations. Because the nominal cadences of KMTNet is 2.25–3.0 hr$^{-1}$, which is sufficient for the $A \lesssim 50$ region, no follow-up observations were taken before HJD′ = 9702.625. Then, the MAP collaboration conducted follow-up observations using the 1.0 m telescopes at three southern sites, CTIO (LCOC), SAAO (LCOS), and SSO (LCOA), of the LCO global network, for which the normal band is the $I$ band. At UT 16:58 on 3 May 2022 (HJD′ = 9703.21), the KMTNet "auto-followup" system substituted the BLG01 observations ($\Gamma = 1.5$ hr$^{-1}$ for KMTS and $\Gamma = 2.0$ hr$^{-1}$ for KMTC) with the BLG41 observations for KMTS and KMTC on the peak. The *μ*FUN group observed the event using a 0.18 m Newtonian telescope at El Sauce Observatory in Chile (CHI-18), for which the filter is similar to the SDSS-*r*′ band.

**Table 1.** Data information with corresponding data reduction method

| Collaboration | Site | Name | Filter | Coverage (HJD′)[1] | $N_{\rm data}$ | Reduction Method | $(k, e_{\rm min})$[2] |
|---|---|---|---|---|---|---|---|
| KMTNet | SSO | KMTA22(*I*) | *I* | 9638.3 – 9749.2 | 183 | pySIS[3] | (1.142, 0.003) |
| KMTNet | SSO | KMTA41(*I*)[6] | *I* | 9638.3 – 9749.2 | 255 | pySIS | (1.505, 0.000) |
| KMTNet | CTIO | KMTC22(*I*) | *I* | 9630.8 – 9750.6 | 547 | pySIS | (0.929, 0.001) |
| KMTNet | CTIO | KMTC41(*I*)[6] | *I* | 9630.9 – 9750.7 | 1010 | pySIS | ... |
| KMTNet | CTIO | KMTC41(*V*) | *V* | 9633.9 – 9749.8 | 100 | pySIS | (1.379, 0.000) |
| KMTNet | SAAO | KMTS22(*I*) | *I* | 9631.6 – 9750.5 | 310 | pySIS | (1.288, 0.000) |
| KMTNet | SAAO | KMTS41(*I*)[6] | *I* | 9631.6 – 9750.5 | 556 | pySIS | (1.205, 0.000) |
| KMTNet | SAAO | KMTS41(*V*) | *V* | 9632.6 – 9750.3 | 57 | pySIS | (2.035, 0.000) |
| MOA | Mt. John Observatory | MOA(Red) | Red | 9621.2 – 9756.8 | 318 | Bond et al. (2001) | (1.059, 0.000) |
| MAP | SSO | LCOA(*I*) | *I* | 9703.0 – 9726.1 | 29 | pySIS | (0.814, 0.002) |
| MAP | CTIO | LCOC(*r*) | *r* | 9703.7 – 9704.8 | 46 | ISIS[4] | (0.994, 0.002) |
| MAP | CTIO | LCOC(*I*) | *I* | 9702.7 – 9711.9 | 36 | pySIS | (1.047, 0.000) |
| MAP | SAAO | LCOS(*I*) | *I* | 9702.6 – 9712.6 | 155 | pySIS | (1.143, 0.001) |
| *μ*FUN | El Sauce Observatory | CHI-18 | 580–700nm | 9703.7 – 9704.8 | 79 | ISIS | (1.152, 0.000) |
| KMTNet | SAAO | | *I* | | 551 | pyDIA[5] | ... |
| KMTNet | SAAO | | *V* | | 57 | pyDIA[5] | ... |

[1] HJD′ = HJD − 2450000
[2] $(k, e_{\rm min})$ are the error rescaling factors as described in Yee et al. (2012).
[3] Albrow et al. (2009)
[4] Alard & Lupton (1998); Alard (2000); Zang et al. (2018)
[5] Albrow (2017)
[6] There is a bleed trail near the source on the *I*-band KMTNet BLG41 images. The photometry of the KMTC41 (*I*) light curve was significantly affected by the bleed trail so we exclude the KMTC41 (*I*) light curve from the analysis. The KMTS41 (*I*) and KMTA41 (*I*) light curves were not affected by the bleed trail due to worse seeing so we include them in the analysis.

At UT 02:48 on 4 May 2022 (HJD′ = 9703.62), W. Zang found that the real-time LCOS data after HJD′ = 9703.56 showed a bump-type anomaly. Then, he replaced the *I*-band observations of LCOC that night with the SDSS-*r* band observations to exclude the potential 1L2S model by the color argument. Following the alert from W. Zang, the KMTNet "auto-followup" system further substituted the BLG02 observations with the BLG41 observations for the KMTC on the peak, and thus the KMTC observations afterward had a cadence of $\Gamma = 7.0$ hr$^{-1}$ that night. However, considering the existence of high-cadence KMTC follow-up observations for this event, the MAP collaboration put a higher priority on another high-magnification event, KMT-2022-BLG-0567, which peaked at the Chile zone (HJD′ = 9703.79), thus LCOC only got 1.3-hr data on the peak of KMT-2022-BLG-0440. Nevertheless, these LCOC data still play an important role in excluding the resonant solutions and the 1L2S model.

Independent of the follow-up observations that responded to the alert of HighMagFinder, the MOA group increased the cadence to ∼ 2.3 hr$^{-1}$ around the peak (9702.9 < HJD′ < 9704.3).

### 2.3 Data Reduction and Error Renormalization

For the light-curve analysis, the data were reduced by the difference image analysis (DIA, Tomaney & Crotts 1996; Alard & Lupton 1998) pipelines of the individual groups: pySIS (Albrow et al. 2009, Yang et al. in prep) for the KMTNet and *I*-band LCO data, Bond et al. (2001) for the MOA data and ISIS (Alard & Lupton 1998; Alard 2000; Zang et al. 2018) for the *r*-band LCO and CHI-18 data. Because the 1L2S analysis requires measuring the source $r - I$ color and calibrating it to $V - I$ by field stars, the *r*-band LCOC data were reduced using a custom ISIS pipeline that simultaneously yields the light curve on the same magnitude system as field-star photometry.

It was not until the DIA data reduction that we found that the KMTNet images have two problems. First, the source is at the edge of both BLG22 and BLG41 images, at which the optical collimation is not perfect, so the point spread function (PSF) is elliptical for many images. The PSF problem is more severe in the KMTC22 (*I*) images, for which some images have double-peak PSF. Nevertheless, the KMTNet light curves from the pySIS pipeline are still of good quality and consistent with the MOA and follow-up data, and the only problem is systematics for the microlensing parallax measurements (Gould 1992, 2000, 2004). Second, on the *I*-band KMTNet BLG41 images the source was near a bleed trail from a saturated star, which has $(\alpha, \delta)_{\rm J2000} = (17:58:20.22, -32:15:56.46)$, and $G = 11.64$ (Gaia Collaboration et al. 2016, 2018). The saturated star is located outside the field of view of the KMTNet BLG22 field. As shown in the lower panel of Figure 1, the KMTC41 (*I*) data are significantly affected by the bleed trail and exhibit different behavior from the KMTC22 (*I*) and LCOC (*r*) data over the peak, while the KMTS41 (*I*) data are consistent with the KMTS22 (*I*) and LCOS (*I*) data. This is because that seeing at SAAO and SSO is worse than at CTIO so the bleed trail is much weaker in the SAAO and SSO images. For the KMTC41 (*I*) images around the peak, the background flux is ∼ 700 ADU/pixel while the spike has a flux of ∼ 1100 ADU/pixel, i.e., a significance level of ∼ 15$\sigma$. For the KMTS41 (*I*) and KMTA41 (*I*) images, the background flux is ∼ 1000 ADU/pixel while the spike has a flux of ∼ 1200 ADU/pixel, with a significance level of only $\sim 6\sigma$. Therefore, we exclude the KMTC41 (*I*) data in the analysis but keep the KMTS41 (*I*) and KMTA41 (*I*) data.

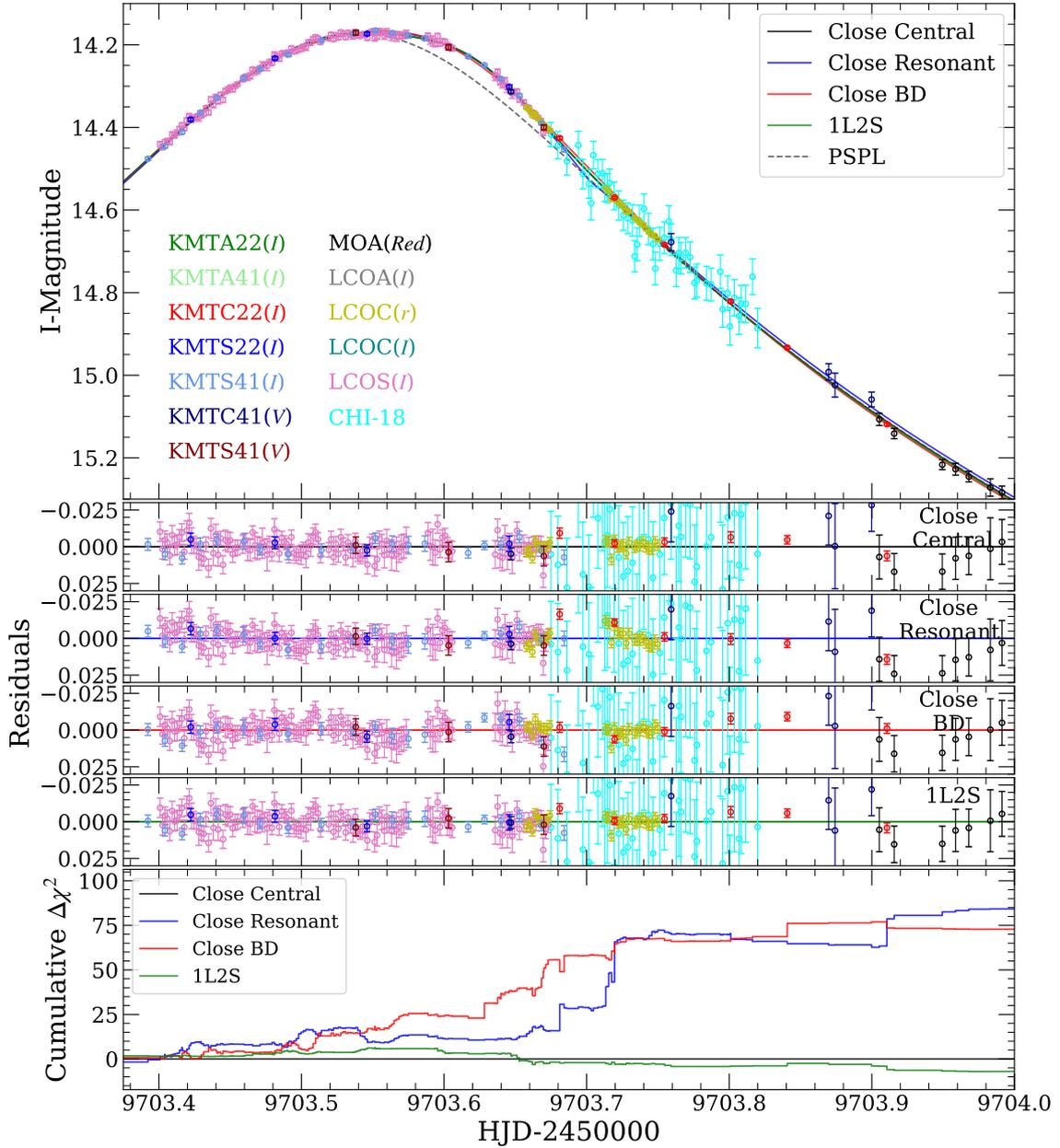

**Figure 2.** A close-up of the anomaly and the residuals to different close ($s < 1$) models. The wide ($s > 1$) models exhibit similar light curves. Different models are shown with different colors, and the corresponding parameters are presented in Table 2. The bottom panel shows the cumulative distribution of $\chi^2$ differences for different models compared to the "Close Central" model. The "resonant" and "BD" solutions are excluded by $\Delta\chi^2 > 70$. The 1L2S model cannot be rejected by the light-curve analysis but is excluded by the color of the second source shown in Section 5.

We conduct pyDIA photometry (Albrow 2017) of the KMTS41 data to measure the source $V - I$ color and locate the source on the color-magnitude diagram (CMD). The *I*-band magnitude reported in this paper has been calibrated to the standard *I*-band magnitude using the OGLE-III star catalog (Szymański et al. 2011). The photometric error bars estimated by the DIA pipelines were re-adjusted using the prescription described in Yee et al. (2012) with the best 2L1S model, which enables $\chi^2$/dof for each data set to become unity, where "dof" is the degree of freedom. We also fit the best 2L1S model with the error bars rescaling parameters free and obtain the same result. In Table 1, we list the basic observational information, data reduction method, and rescaling factors of each data set.

## 3 BINARY-LENS SINGLE-SOURCE ANALYSIS

Figure 2 shows an enlargement of the peak region together with different models. The observed data exhibit a short-lived ($\sim 4$ hr) bump after the peak of an otherwise Paczyński (point-source point-lens, hereafter PSPL) curve (Paczyński 1986), which is described by three parameters ($t_0$, $u_0$, $t_E$), i.e., the time of lens-source closest approach, $t_0$, the impact parameter of this approach, $u_0$ (in units of the angular Einstein radius $\theta_E$), and the Einstein radius crossing time,

$$t_E = \frac{\theta_E}{\mu_{\rm rel}}; \qquad \theta_E = \sqrt{\kappa M_L \pi_{\rm rel}}, \tag{1}$$

where $\kappa \equiv \frac{4G}{c^2 {\rm au}} \simeq 8.144 \frac{\rm mas}{M_\odot}$, $M_L$ is the lens mass, and $(\pi_{\rm rel}, \mu_{\rm rel})$ are the lens-source relative (parallax, proper motion). The bump was supported by the KMTS, LCOS, KMTC, and LCOC (r) data, so the anomaly is secure. The CHI-18 data captured the end of the bump, but due to the large photometric uncertainty the CHI-18 data do not contribute to the identification of the anomaly.

For a static 2L1S model, three additional parameters $(q, s, \alpha)$ define the binary geometry: the mass ratio between the secondary lens and the primary lens, $q$, the projected separation between the binary lenses normalized to $\theta_E$, $s$, and the angle between the source trajectory and the binary axis, $\alpha$. The seventh parameter, $\rho$, is the angular source radius $\theta_*$ scaled to $\theta_E$, i.e., $\rho = \theta_*/\theta_E$. In addition, because the source flux could be blended with other unlensed stars and the lens flux, we introduce two linear parameters $(f_{S,i}, f_{B,i})$ for each data set $i$ to represent the source flux and any blended flux. During our fitting process, $f_{S,i}$ and $f_{B,i}$ are free for all of the models, with a uniform prior in the flux scale. We employ the advanced contour integration code (Bozza 2010; Bozza et al. 2018), VBBinaryLensing, to calculate the 2L1S magnification $A(t)|_{(t_0, u_0, t_E, \rho, q, s, \alpha)}$.

To explore the 2L1S parameter space and locate all the local $\chi^2$ minima, we first conduct a sparse grid search covering a wide range of parameters over $(\log s, \log q, \log \rho, \alpha)$ and then conduct a denser grid search in the subspace that contains the local minima. The sparse grid contains 61 values evenly distributed in $-1.5 \leqslant \log s \leqslant 1.5$, 61 values evenly distributed in $-6 \leqslant \log q \leqslant 0$, nine values evenly distributed in $-4.0 \leqslant \log \rho \leqslant -1.6$, and 12 initial values evenly distributed in $0° \leqslant \alpha < 360°$. The initial values of $(t_0, u_0, t_E)$ are seeded at the PSPL fitting values, where $t_0({\rm HJD}') = 9703.54$, $u_0 = 0.004$, $t_E = 41$ days. We search for the minimum $\chi^2$ by Markov chain Monte Carlo (MCMC) $\chi^2$ minimization using the emcee ensemble sampler (Foreman-Mackey et al. 2013), during which $(\log s, \log q, \log \rho)$ are held fixed while $(t_0, u_0, t_E, \alpha)$ are allowed to vary.

The upper panel of Figure 3 shows the $\chi^2$ surface in the $(\log s, \log q)$ plane from the sparse grid search. We identify two local minima with a brown dwarf (BD) mass ratio $(\log q \sim -1.5)$ and so label them as "Close BD" and "Wide BD". In addition, there are distinct minima within $-0.25 \leqslant \log s \leqslant 0.25$, $-5.0 \leqslant \log q \leqslant -3.5$ and $-4.0 \leqslant \log \rho \leqslant -2.8$. We note that the topology of this result is similar to the topology of KMT-2021-BLG-1689 (Yang et al. 2022), for which a denser grid search for the lower-$q$ minima revealed the "central-resonant" caustic degeneracy. Therefore, we conduct a denser grid search, which contains 251 values evenly distributed in $-0.25 \leqslant \log s \leqslant 0.25$, 31 values evenly distributed in $-5.0 \leqslant \log q \leqslant -3.5$, five values evenly distributed in $-4.0 \leqslant \log \rho \leqslant -2.8$, and 12 initial values evenly distributed in $0° \leqslant \alpha < 360°$. The resulting projected $\chi^2$ distribution is shown in the lower panel of Figure 3. We find four distinct local minima, and two of them have central caustics while the other two have resonant caustics. We thus label these four solutions as "Close Central", "Wide Central", "Close Resonant", and "Wide Resonant".

We then refine the best-fit solutions by MCMC with all seven parameters of the static 2L1S model free and then further explore the minimum $\chi^2$ by a downhill[2] approach, and Table 2 presents the

---

[2] We use a function based on the Nelder-Mead simplex algorithm from the SciPy package. See https://docs.scipy.org/doc/scipy/reference/generated/scipy.optimize.fmin.html#scipy.optimize.fmin

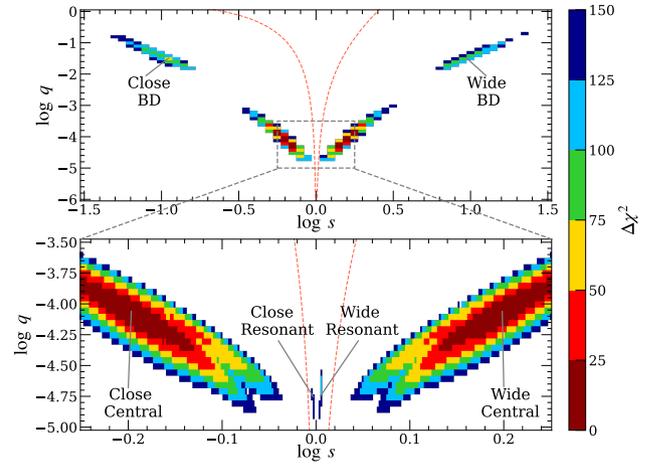

**Figure 3.** $\chi^2$ surface in the $(\log s, \log q)$ plane from the grid search. The upper panel shows the space that is evenly divided on a $(61 \times 61)$ grid with ranges of $-1.5 \leqslant \log s \leqslant 1.5$ and $-6.0 \leqslant \log q \leqslant 0$, respectively. The lower panel displays the space on a $(251 \times 31)$ grid with ranges of $-0.25 \leqslant \log s \leqslant 0.25$ and $-5.0 \leqslant \log q \leqslant -3.5$, respectively. The labels "Close BD", "Wide BD", "Close Central", "Wide Central", "Close Resonant", and "Wide Resonant" indicate six distinct minima. The two red dashed lines indicate the boundaries between resonant and non-resonant caustics using Equation (59) of Dominik (1999). Grid points with $\Delta \chi^2 > 150$ are marked as blank.

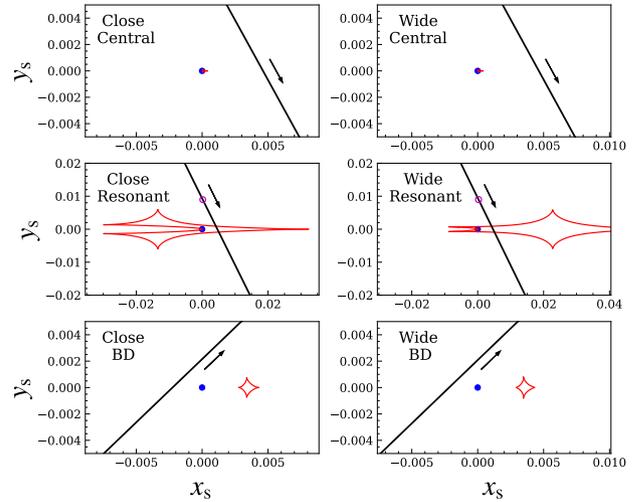

**Figure 4.** Caustic geometries of six 2L1S solutions. In each panel, the red lines show the caustic, the blue dot indicates the location of the host star, the black line represents the source-lens relative trajectory, and the line with an arrow indicates the direction of the source motion. The "Close Resonant", and "Wide Resonant" solutions show caustic-crossing features so $\rho$ is constrained at the $> 3\sigma$ level and the radii of the magenta circles represent the source radii.

resulting parameters and the source and blend brightness in the standard $I$ and $V$ band. Figure 4 shows the caustic geometries of the six solutions. For the two "Resonant" solutions, the anomaly was caused by a caustic crossing, so $\rho$ is well constrained for these two solutions. For the other four solutions, the bump was the result of a cusp approach, so finite-source effects are only marginally detected and the observed data are consistent with a point-source model within $1\sigma$ level. We also note that for the two "BD" solutions, $q$ and $s$ for the close/wide degeneracy are consistent with Equation (5.21) of An

**Table 2.** 2L1S static models for KMT-2022-BLG-0440

| Parameters | Close | | | Wide | | |
|---|---|---|---|---|---|---|
| | Central | Resonant | BD | Central | Resonant | BD |
| $\chi^2$/dof | 2637.9/2638 | 2737.7/2638 | 2712.6/2638 | 2638.3/2638 | 2729.0/2638 | 2712.7/2638 |
| $t_0 - 9703$ (HJD') | 0.5395 ± 0.0004 | 0.5423 ± 0.0002 | 0.4353 ± 0.0089 | 0.5395 ± 0.0004 | 0.5424 ± 0.0002 | 0.4334 ± 0.0085 |
| $u_0 (10^{-3})$ | 4.1 ± 0.1 | 4.2 ± 0.1 | 1.5 ± 0.2 | 4.1 ± 0.1 | 4.2 ± 0.2 | 1.5 ± 0.1 |
| $t_E$ (days) | 41.08 ± 1.37 | 40.03 ± 1.35 | 41.97 ± 1.54 | 41.29 ± 1.41 | 40.06 ± 1.45 | 42.62 ± 1.53 |
| $\rho (10^{-3})$ | < 1.032 | 0.881 ± 0.055 | < 0.891 | < 1.028 | 0.904 ± 0.098 | < 0.803 |
| $\alpha$ (degree) | 298.83 ± 0.43 | 296.34 ± 0.28 | 43.89 ± 1.05 | 298.77 ± 0.41 | 296.03 ± 0.51 | 43.89 ± 0.97 |
| $s$ | 0.6359 ± 0.0195 | 0.9933 ± 0.0004 | 0.1112 ± 0.0075 | 1.5799 ± 0.0463 | 1.0114 ± 0.0004 | 9.3771 ± 0.6409 |
| $q (10^{-4})$ | 0.880 ± 0.123 | 0.176 ± 0.009 | 325.0 ± 48.1 | 0.876 ± 0.116 | 0.180 ± 0.009 | 339.7 ± 48.9 |
| $\log q$ | −4.060 ± 0.061 | −4.754 ± 0.023 | −1.492 ± 0.061 | −4.061 ± 0.058 | −4.744 ± 0.023 | −1.473 ± 0.061 |
| $I_{S,OGLE}$ | 20.172 ± 0.037 | 20.142 ± 0.037 | 20.196 ± 0.041 | 20.167 ± 0.037 | 20.139 ± 0.039 | 20.174 ± 0.040 |
| $V_{S,OGLE}$ | 21.983 ± 0.037 | 21.951 ± 0.037 | 22.007 ± 0.041 | 21.988 ± 0.038 | 21.953 ± 0.040 | 22.006 ± 0.040 |
| $I_{B,OGLE}$ | 18.236 ± 0.052 | 18.241 ± 0.052 | 18.232 ± 0.052 | 18.235 ± 0.052 | 18.241 ± 0.052 | 18.232 ± 0.052 |
| $V_{B,OGLE}$ | 19.999 ± 0.103 | 20.004 ± 0.104 | 19.995 ± 0.103 | 19.998 ± 0.103 | 20.004 ± 0.104 | 19.995 ± 0.103 |

NOTE. HJD' = HJD − 2450000. The upper limit on $\rho$ is $3\sigma$. $t_0$ and $u_0$ take the position of the host star as the origin.

(2005) within $1\sigma$. The "Close Central" solution provides the best fit to the data, and the "Wide Central", "Close BD", "Wide BD", "Close Resonant", and "Wide Resonant" solutions are disfavored by $\Delta\chi^2 = 0.4, 74.7, 74.8, 99.8,$ and 91.1, respectively. Figure 2 displays the residuals of the three "Close" solutions and the cumulative distribution of $\chi^2$ differences for the "Close BD" and "Close Resonant" solutions compared to the "Close Central" solution. We find that the "Close Resonant" solution cannot well fit the end of the short-lived bump and both the KMTC22 (*I*) and LCOC (*r*) data contribute to the main $\chi^2$ differences. For the "Close BD" solution, the KMTS41 (*I*) and LCOS (*I*) data both provide a worse fit. The "Wide" solutions exhibit the same feature, so we exclude the "Close BD", "Wide BD", "Close Resonant" and "Wide Resonant" solutions and only investigate the two "Central" solutions in the following analysis. In addition, we also fit the 2L1S model without the KMTS41 (*I*) and KMTA41 (*I*) data and find that the "resonant" and "BD" solutions can be excluded by $\Delta\chi^2 > 40$.

Because the event is not short, we also check whether the fit can be improved by the annual microlensing parallax effect, although the KMTNet survey data may have systematics due to an elliptical PSF and a bleed trail (see Section 2). Indeed, we find systematics in the KMTNet data. The resulting parallax value is ∼ 1, which is of very low probability though not impossible (e.g., Ryu et al. 2019). In addition, different KMTNet data sets show inconsistent cumulative distributions for $\Delta\chi^2$ between the static and parallax models. Therefore, we conclude that the annual microlensing parallax effect cannot be reliably constrained and adopt the static models as our final result.

## 4 COLOR-MAGNITUDE DIAGRAM (CMD)

Before presenting the 1L2S analysis, we conduct the CMD analysis here for two reasons. First, the CMD analysis can yield the angular source radius $\theta_*$ (Yoo et al. 2004), and so

$$\theta_E = \frac{\theta_*}{\rho}, \quad \mu_{rel} = \frac{\theta_E}{t_E}. \tag{2}$$

For a 1L2S solution with very small $\mu_{rel}$, one can make a kinematic argument that the solution is unlikely. Following the $\mu_{rel}$ distribution of observed planetary microlensing events (Gould 2022), Jung et al.

**Table 3.** CMD parameters, source properties and derived $\theta_E$ and $\mu_{rel}$ for KMT-2022-BLG-0440

| $(V - I)_{RC}$ | 2.143 ± 0.007 |
|---|---|
| $I_{RC}$ | 15.920 ± 0.022 |
| $(V - I)_{RC,0}$ | 1.06 ± 0.03 |
| $I_{RC,0}$ | 14.53 ± 0.04 |
| $(V - I)_S$ | 1.820 ± 0.003 |
| $I_S$ | 20.17 ± 0.04 |
| $(V - I)_{S,0}$ | 0.73 ± 0.03 |
| $I_{S,0}$ | 18.77 ± 0.06 |
| $\theta_*$ ($\mu$as) | 0.579 ± 0.026 |
| $\theta_E$ (mas) | > 0.56 |
| $\mu_{rel}$ (mas yr$^{-1}$) | > 5.0 |

The lower limits on $\theta_E$ and $\mu_{rel}$ are $3\sigma$.

(2023) showed that the fraction of events with proper motions lower than a given $\mu_{rel} \ll \sigma_\mu$ is

$$p(\leqslant \mu_{rel}) \to \frac{\mu_{rel}^2}{4\sigma_\mu^2} \to 2.8 \times 10^{-2} \left(\frac{\mu_{rel}}{1 \text{ mas yr}^{-1}}\right)^2, \tag{3}$$

where we adopt $\sigma_\mu = 3.0$ mas yr$^{-1}$ for the proper motion dispersion lenses. Second, the CMD analysis can yield the intrinsic color and thus make a color argument for (or against) the 1L2S model (Gaudi 1998).

Figure 5 shows the $V - I$ versus $I$ CMD for KMT-2022-BLG-0440, which is constructed using the OGLE-III catalog stars (Szymański et al. 2011) within 2.5' centered on the source position. The centroid of the red clump (RC) is at $(V - I, I)_{RC} = (2.143 \pm 0.007, 15.920 \pm 0.022)$, and the intrinsic color and de-reddened brightness of the red clump are $(V - I, I)_{RC,0} = (1.06 \pm 0.03, 14.53 \pm 0.04)$ (Bensby et al. 2013; Nataf et al. 2013). Thus, the extinction toward this direction for the red clump is $A_I = 1.39 \pm 0.04$ and $E(V - I) = 1.08 \pm 0.03$. For the source color, which is independent of any model, we begin by regression of KMTS41 pyDIA $V$ versus $I$ flux as the lensing magnification changes and obtain $(V - I)_{S,KMT} = 2.123 \pm 0.002$.

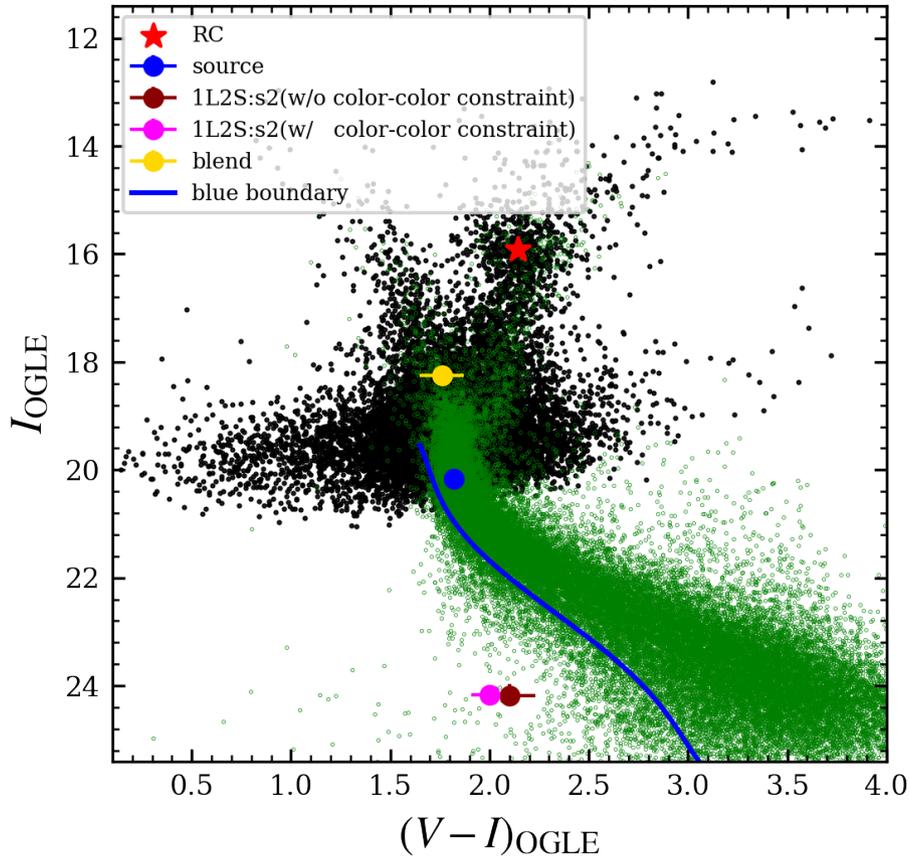

**Figure 5.** Color-magnitude diagram (CMD) of KMT-2022-BLG-0440, which is constructed by the field stars (black points) within 2.5′ centered on the source position using the OGLE-III star catalog (Szymański et al. 2011). The red asterisk represents the centroid of the red clump (RC), the blue dot indicates the source of the 2L1S model, and the yellow dot shows the blended light of the OGLE-III baseline object. The magenta and brown dots represent the second source of the 1L2S models with and without the color-color constraint (i.e., the third and first solutions in Table 4). The green points represent the HST CMD of Holtzman et al. (1998) whose RC has been matched to that of OGLE-III. The blue line indicates the blue boundary of the bulge main-sequence stars derived from stellar isochrones with [M/H] = −1.0 and age > 9 Gyr.

Then, we calibrate the color to the OGLE-III magnitude scale using the bright field stars of the KMTS41 pyDIA and the OGLE-III star catalogs and obtain $(V - I)_S = 1.820 \pm 0.003$. According to Figure 5, the source is a typical star in the Galactic bulge. From the Bayesian analysis in Section 6, the source distance is $8.7^{+1.1}_{-1.0}$ kpc. Following the procedure of Yang et al. (2022), we find that the extinction for the source star is $A_I = 1.40 \pm 0.04$ and $E(V - I) = 1.09 \pm 0.03$, leading to the source intrinsic color of $(V - I)_{S,0} = 0.73 \pm 0.03$ and the de-reddened source brightness of $I_{S,0} = 18.77 \pm 0.06$. Using the color/surface-brightness (CSB) relation of Adams et al. (2018),

$$\log(2\theta_*) = 0.378(V - I)_{S,0} + 0.542 - 0.2 I_{S,0}, \qquad (4)$$

we obtain the angular source radius $\theta_* = 0.579 \pm 0.026\ \mu$as. In Table 3, we summarize the CMD parameters and the source values.

The OGLE-III baseline object has $(V - I, I)_{\rm base} = (1.771 \pm 0.099, 18.066 \pm 0.044)$, yielding a blend of $(V - I, I)_B = (1.76 \pm 0.11, 18.24 \pm 0.05)$. We display the blend in Figure 5, which shows that the blend probably belongs to the foreground main-sequence branch. We also check the astrometric alignment of the baseline object by four $i'$-band baseline images of the 3.6m Canada-France-Hawaii Telescope (CFHT). These images were taken in 2018 with seeing FWHM of 0.″60–0.″70. We find that the astrometric offset is $\Delta\theta(N, E) = (136, 393)$ mas, so the majority of the blended light is not the lens light.

## 5 SINGLE-LENS BINARY-SOURCE ANALYSIS

If the second source is much fainter and passes closer to the lens, a 1L2S event can also produce a short-lived bump, which is similar to planetary anomalies (Gaudi 1998). There have been several cases with plausible planetary anomalies that proved to be caused by binary sources (Hwang et al. 2013; Jung et al. 2017; Rota et al. 2021; Han et al. 2022c) or even triple sources (Hwang et al. 2018). We thus investigate whether a 1L2S model can account for the observed anomaly.

### 5.1 Light-curve Analysis

For a static 1L2S model, the light curve is the superposition of the 1L1S curves of two sources, thus the total effective magnification changes over time at a certain waveband $\lambda$, $A_\lambda(t)$, can be expressed as (Hwang et al. 2013)

$$A_\lambda(t) = \frac{A_1(t) f_{1,\lambda} + A_2(t) f_{2,\lambda}}{f_{1,\lambda} + f_{2,\lambda}} = \frac{A_1(t) + q_{f,\lambda} A_2(t)}{1 + q_{f,\lambda}}, \qquad (5)$$

Table 4. 1L2S models for KMT-2022-BLG-0440

| | | | | |
|---|---|---|---|---|
| color argument? | × | √ | × | √ |
| color-color constraint? | × | × | √ | √ |
| $\chi^2_{\text{total}}$/dof | 2627.0/2635 | 2646.2/2635 | 2628.9/2635 | 2674.5/2635 |
| $t_{0,1} - 9703$ (HJD′) | 0.5351 ± 0.0008 | 0.5356 ± 0.0007 | 0.5349 ± 0.0008 | 0.5364 ± 0.0007 |
| $t_{0,2} - 9703$ (HJD′) | 0.6235 ± 0.0013 | 0.6240 ± 0.0012 | 0.6234 ± 0.0013 | 0.6254 ± 0.0011 |
| $u_{0,1}(10^{-3})$ | 4.1 ± 0.2 | 4.1 ± 0.2 | 4.1 ± 0.2 | 4.1 ± 0.2 |
| $u_{0,2}(10^{-3})$ | 1.4 ± 0.3 | 1.3 ± 0.5 | 1.5 ± 0.2 | 0.7 ± 0.6 |
| $t_E$ (days) | 42.05 ± 1.52 | 41.38 ± 1.35 | 41.97 ± 1.49 | 41.80 ± 1.51 |
| $\rho_1 (10^{-3})$ | < 2.786 | < 2.661 | < 3.112 | < 2.483 |
| $\rho_2 (10^{-3})$ | < 2.075 | < 2.017 | < 2.051 | 1.454 ± 0.286 |
| $q_{f,I}$ | 0.027 ± 0.005 | 0.023 ± 0.004 | 0.027 ± 0.005 | 0.014 ± 0.003 |
| $q_{f,V}$ | 0.020 ± 0.005 | 0.008 ± 0.002 | 0.023 ± 0.005 | 0.005 ± 0.001 |
| $q_{f,r}$ | 0.027 ± 0.006 | 0.024 ± 0.005 | 0.026 ± 0.005 | 0.008 ± 0.002 |
| $I_{S1,\text{OGLE}}$ | 20.222 ± 0.041 | 20.203 ± 0.039 | 20.221 ± 0.041 | 20.207 ± 0.047 |
| $V_{S1,\text{OGLE}}$ | 22.021 ± 0.041 | 21.979 ± 0.040 | 22.025 ± 0.041 | 21.993 ± 0.054 |
| $I_{S2,\text{OGLE}}$ | 24.185 ± 0.213 | 24.303 ± 0.200 | 24.159 ± 0.182 | 24.912 ± 0.257 |
| $V_{S2,\text{OGLE}}$ | 26.288 ± 0.269 | 27.203 ± 0.259 | 26.156 ± 0.215 | 27.845 ± 0.323 |
| $I_{B,\text{OGLE}}$ | 18.231 ± 0.052 | 18.234 ± 0.052 | 18.231 ± 0.052 | 18.231 ± 0.052 |
| $V_{B,\text{OGLE}}$ | 19.996 ± 0.103 | 20.001 ± 0.103 | 19.996 ± 0.103 | 19.998 ± 0.103 |
| $\chi^2_{\text{color}}$ | ... | ... | 0.12 | 0.04 |
| $(V-I)_{S2}$ | 2.10 ± 0.13 | 2.90 ± 0.09 | 2.00 ± 0.09 | 2.93 ± 0.12 |

NOTE. HJD′ = HJD − 2450000. The upper limits on $\rho_1$ and $\rho_2$ are 3$\sigma$.

$$q_{f,\lambda} = \frac{f_{2,\lambda}}{f_{1,\lambda}}, \quad (6)$$

where $f_{j,\lambda}$ and $A_j(t)$ ($j = 1, 2$) respectively represent the flux at waveband $\lambda$ and magnification of each source, and $q_{f,\lambda}$ is therefore the flux ratio between the primary and the second sources at waveband $\lambda$. We use $q_{f,I}$ for the KMT $I$-band data, LCO $I$-band data and MOA data, $q_{f,V}$ for the KMT $V$-band data, and $q_{f,r}$ for the LCO $r$-band data and CHI-18 data. We also introduce parameters ($f_{S,i}$, $f_{B,i}$) for each data set $i$ at waveband $\lambda$ to represent the total flux of the two sources and any blended flux. Then, the observed flux $f_i(t)$ at waveband $\lambda$ is modeled as

$$f_i(t) = f_{S,i} \times A_\lambda(t) + f_{B,i}. \quad (7)$$

We search for the 1L2S model by MCMC. During the fitting process, $f_{S,i}$ and $f_{B,i}$ are free, with a uniform prior in the flux scale. The 1L2S parameters together with the source and blend brightness in the standard $I$ and $V$ band are listed in Table 4 as while as the 1L2S model light curve is shown in Figure 2. We find that the best-fit 1L2S model is favored by $\Delta\chi^2 = 10.9$ compared to the best-fit 2L1S model (i.e., the 2L1S "Close Central" solution). Therefore, we cannot rule out the 1L2S model by the light-curve analysis alone. However, a physically reasonable 1L2S model should follow other constraints, and below we add these constraints to test the 1L2S model.

Because the two sources both can yield the measurement of $\theta_E$, a 1L2S model must satisfy

$$\theta_E = \frac{\theta_{*,1}}{\rho_1} = \frac{\theta_{*,2}}{\rho_2}; \quad \frac{\rho_2}{\rho_1} = \frac{\theta_{*,2}}{\theta_{*,1}}. \quad (8)$$

We add a $\chi^2_{\theta_E}$ into the $\chi^2_{\text{total}}$ during the MCMC process,

$$\chi^2_{\theta_E} = \left(\frac{\rho_2}{\rho_1} - \frac{\theta_{*,2}}{\theta_{*,1}}\right)^2 \bigg/ \left(\sigma_{\text{CSB}} \frac{\theta_{*,2}}{\theta_{*,1}}\right)^2, \quad (9)$$

where we adopt $\sigma_{\text{CSB}} = 1\%$ to account for the uncertainty of $\frac{\theta_{*,2}}{\theta_{*,1}}$ from the CSB relation. The typical uncertainty of the Adams et al. (2018) CSB relation is 5%, but because the two source stars must come from the same gas cloud and we use the same CSB relation to calculate $\theta_{*,1}$ and $\theta_{*,2}$, the uncertainty of $\frac{\theta_{*,2}}{\theta_{*,1}}$ is much smaller than 5%. We also test the results with $\sigma_{\text{CSB}} = 10\%$, which has almost no influence on the conclusions below. We also consider the "kinematic argument" and restrict $\mu_{\text{rel}} > 1$ mas yr$^{-1}$ during the MCMC process. However, the two constraints only increase the $\chi^2_{\text{total}}$ by 7.2. This is because the light curves only provide a weak constraint on $\rho_1$ and $\rho_2$, and for the primary and second sources the 1L2S model is consistent with a point-source model by < 1$\sigma$ and < 3$\sigma$, respectively. Therefore, Equation (8) and "kinematic argument" can be easily satisfied. We thus check whether the 1L2S model can be excluded by the "color argument".

### 5.2 "Color Argument"

Mao & Paczynski (1991) first pointed out that different source colors will make the light curve of 1L2S events color-dependent. Gaudi (1998) proposed that the 1L2S and 2L1S models can be distinguished by the color difference expected for the two sources of different luminosities. For the present case, there are three KMTS41 ($V$) points, one KMTC41 ($V$) point, and nine LCOC ($r$) points taken during the anomaly. We begin by only using the KMTNet $V$-band data for the "color argument", to further investigate the role of the LCOC ($r$) data.

The color of the primary source is well determined by the data outside the anomaly region, as shown in Section 4. For the second source, the 1L2S modeling yields $(V − I, I)_{S2} = (2.10 ± 0.13, 24.18 ± 0.21)$. We locate the second source in the OGLE-III CMD and calibrate the CMD of Holtzman et al. (1998) HST observations to the OGLE-III CMD using their positions of red clump (Bennett et al. 2008). As shown in Figure 5, the second source is inconsistent with the bulge main-sequence stars, but for a physically reasonable 1L2S model, the second source should be a bulge main-sequence star. To quantify the discrepancy, we restrict the second source to the bulge main-sequence stars. However, due to the HST photometric errors, which are typically $\sigma(V-I) \sim 0.15$ for the blue HST field stars at $I \sim 24.2$, the actual blue boundary of the bulge main-sequence stars is redder than what Figure 5 shows. According to the spectroscopic observations for bulge red clump stars (Rojas-Arriagada et al. 2017; Zoccali et al. 2017), main-sequence, turn-off, and subgiant stars (Bensby et al. 2017), > 99% of the bulge stars have the metallicity of [Fe/H] > $-1.25$. For the metal-poor population ([Fe/H] < $-0.8$), Bensby et al. (2017) found that the stellar ages are $\geqslant 9$ Gyr and the [$\alpha$/Fe] ratio is about 0.2–0.4. Therefore, we estimate the blue boundary of the bulge main-sequence stars using stellar isochrones of Bressan et al. (2012), with [M/H] = $-1.0$ and age > 9 Gyr. For $3.5 \leqslant M_I \leqslant 10$, we find that the blue boundary of the selected stellar isochrones can be well fitted by a quintic function of

$$(V - I) = 0.00117(M_I - 7)^5 + 0.00292(M_I - 7)^4 - 0.0215(M_I - 7)^3 \\ - 0.0267(M_I - 7)^2 + 0.355(M_I - 7) + 1.39, \quad (10)$$

where $M_I$ is the absolute magnitude in the $I$ band. We show this relation as the blue boundary in Figure 5. During the MCMC process, we reject the solutions whose $(V-I)_{S,2}$ is bluer than the blue boundaries. The resulting parameters are presented in the third column of Table 4. We find that adding the "color argument" increases the $\chi^2_{\rm total}$ by 19.2, but the 1L2S model is still only disfavored by $\Delta\chi^2 = 8.3$ compared to the best-fit 2L1S model. Therefore, the 1L2S model cannot be ruled out by the "color argument" from the KMTNet $V$-band data.

We now use the color information from the LCOC ($r$) data by converting the $r - I$ color to $V - I$ color. We derive a $V - I$ versus $r - I$ color-color relation by matching the OGLE-III catalog stars and the LCO $r$-band field stars within a 2' square centered on the source position and obtain

$$(V - I)_{\rm LCO} = (2.2656 \pm 0.0033) + \\ (1.7928 \pm 0.0176)(r_{\rm LCO} - I_{\rm OGLE} - 1.40), \quad (11)$$

where 1.40 is a pivot parameter chosen to minimize the covariance between the two linear parameters. To combine the two $V - I$ colors for the second source from the LCO and KMTNet data and consider the uncertainty of $(V - I)_{\rm LCO}$, during MCMC we include a color-color constraint by adding $\chi^2_{\rm color}$ into the total $\chi^2_{\rm total}$,

$$\chi^2_{\rm color} = \frac{[(V - I)_{S2,\rm KMT} - (V - I)_{S2,\rm LCO}]^2}{\sigma^2_{\rm cc}}, \quad (12)$$

where $(V - I)_{S2,\rm KMT}$ is the color from the KMTS41 data, and $\sigma_{\rm cc}$ is the uncertainty of $(V - I)_{S2,\rm LCO}$ from Equation (11). Here the KMTNet $I$- and $V$-band data have been calibrated to the OGLE-III magnitude system. We find that the inclusion of a color-color constraint additionally increases the $\chi^2_{\rm total}$ by 28.3 and the 1L2S model is now disfavored by $\Delta\chi^2 = 36.6$, with $\chi^2_{\rm color} = 0.04$. Table 4 also provides the results with the color-color constraint but without the "color argument". The color-color constraint further constrains the second source color to $(V - I, I)_{S2} = (2.00\pm0.09)$, which is bluer and has a smaller uncertainty compared to the second source color derived only using the KMTNet $V$-band data, then the discrepancy between the second source and the bulge main-sequence stars is $9\sigma$ now. Therefore, the color information from the $V$-band data and the $r$-band data both favor the 2L1S model. We also note that for different

2L1S and 1L2S models, the blend values are positive and almost the same. It is because the second source of the 1L2S model is about 6 magnitude fainter than the blend, thus the second source almost has no effect on the blend.

In addition, $\rho_2$ is constrained to $(1.454 \pm 0.286) \times 10^{-3}$ with the color-color constraint but $\rho_1 < 2.483\times10^{-3}$ at $3\sigma$ level. Because the angular radius of the second source is $\sim 1/3$ of the primary source, this 1L2S model is also disfavored by the $\theta_E$ measurements from the radii of the two sources. We add $\chi^2_{\theta_E}$ into the $\chi^2_{\rm total}$ and find a $\chi^2_{\rm total}$ increment of 18.1, with whose inclusion the 1L2S model is then disfavored by $\Delta\chi^2 = 54.7$ in total compared to the 2L1S model. We find that $\chi^2_{\rm color} = 0.10$ and $\chi^2_{\theta_E} = 0.12$, so the physically reasonable 1L2S model itself contributes to almost all of the $\Delta\chi^2$. We note that $\Delta\chi^2 = 54.7$ is significant enough to exclude the 1L2S model compared to several well-known microlensing studies. For example, Beaulieu et al. (2006) rejected the 1L2S model using $\Delta\chi^2 = 46$ for one of the first microlensing $q < 10^{-4}$ planets, OGLE-2005-BLG-390Lb, and Suzuki et al. (2016) excluded the 1L2S model of the planetary event MOA-2010-BLG-353 (Rattenbury et al. 2015) with $\Delta\chi^2 = 20$ and included this planet in its statistical sample. We also perform the model selection using Akaike's Information Criterion AIC = $\chi^2 + 2n_{\rm param}$ and the Bayesian Information Criterion BIC = $\chi^2 + n_{\rm param} \ln N_{\rm data}$. We find that the 2L1S model has smaller values for both criteria, with $\Delta$ AIC = 61 and $\Delta$ BIC = 78. Hence, we rule out the 1L2S interpretation of KMT-2022-BLG-0440.

In addition, without the KMTS41 ($I$) and KMTA41 ($I$) data, we also fit the 1L2S model including the "color argument", the color-color constraint, and $\chi^2_{\theta_E}$. We find that the 1L2S model is still disfavored by $\Delta\chi^2 > 25$.

## 6 PHYSICAL PARAMETERS

The lens-source relative parallax is given by,

$$\pi_{\rm rel} = \frac{\rm au}{D_L} - \frac{\rm au}{D_S} = \pi_E \theta_E, \quad (13)$$

where $\pi_E$ is the microlensing parallax, $D_L$ and $D_S$ is the lens and the source distances, respectively. If both $\theta_E$ and $\pi_E$ are measured, then Equations (1) and (13) uniquely determine $M_L$ and $\pi_{\rm rel}$, in which case the lens distance, $D_L = {\rm au}/(\pi_{\rm rel}+\pi_S)$, can also be inferred based on an estimate of the source parallax, $\pi_S$. However, as shown in Section 3, neither of the two observables has been unambiguously measured, so we conduct a Bayesian analysis using the Galactic model as priors to estimate the physical parameters of the lens.

The Galactic model mainly consists of three aspects: the lens mass distribution, the stellar number density profile of the lens and the source, and the dynamical distributions of the lens and the source. For the lens mass distribution, we use the initial mass function (IMF) of Kroupa (2001) with a $1.3M_\odot$ and $1.1M_\odot$ cutoff for the disk and the bulge lenses, respectively (Zhu et al. 2017). For the stellar number density profile, we choose the models used by Yang et al. (2021). For the bulge dynamical distributions, we adopt the same model used by Zhu et al. (2017), and for the disk velocity distributions, we use the "Model C" described in Yang et al. (2021), which is based on the `galpy` module (Bovy 2015).

We simulated a sample of $10^7$ events, and for each simulated event $i$ with parameters $t_{E,i}$, $\mu_{{\rm rel},i}$, and $\theta_{E,i}$, we weight it by

$$w_i = \Gamma_i \times p(t_{E,i}) p(\theta_{E,i}), \quad (14)$$

where $\Gamma_i = \theta_{E,i} \times \mu_{{\rm rel},i}$ is the microlensing event rate, $p(t_{E,i})$ represents the probability of $t_{E,i}$ given the error distributions from the $P_{\rm disk}$ is the probability of a lens in the Galactic disk.

**Table 5.** Physical parameters from Bayesian analysis for KMT-2022-BLG-0440

| Solution | Physical Properties | | | | | |
|---|---|---|---|---|---|---|
| | $M_{\rm host}$ ($M_\odot$) | $M_{\rm planet}$ ($M_\oplus$) | $D_{\rm L}$ (kpc) | $r_\perp$ (au) | $\mu_{\rm hel,rel}$ (mas yr$^{-1}$) | $P_{\rm disk}$ |
| Close Central | $0.53^{+0.31}_{-0.26}$ | $15.4^{+9.6}_{-7.4}$ | $3.5^{+1.6}_{-1.6}$ | $1.9^{+0.6}_{-0.7}$ | $7.6^{+1.9}_{-1.4}$ | 97.1% |
| Wide Central | $0.53^{+0.30}_{-0.25}$ | $15.3^{+9.2}_{-7.3}$ | $3.5^{+1.6}_{-1.5}$ | $4.6^{+1.4}_{-1.7}$ | $7.7^{+1.9}_{-1.4}$ | 97.5% |

MCMC as shown in Table 2, and $p(\theta_{{\rm E},i})$ represents the probability of $\theta_{{\rm E},i}$. For $p(\theta_{{\rm E},i})$, we first calculate $\rho_i = \theta_*/\theta_{{\rm E},i}$ and then find the corresponding $\chi^2(\rho_i)$ from the lower envelope of the ($\chi^2$ vs. $\rho$) diagram, which is derived from the MCMC chain. Hence,

$$p(\theta_{{\rm E},i}) = \exp\left[(\chi^2_{\min} - \chi^2(\rho_i))/2\right], \quad (15)$$

where $\chi^2_{\min}$ is the minimum $\chi^2$ of the MCMC chain. As shown in Section 4, the majority of the blended light is not the lens light, so we adopt 50% of the blended light as the upper limit of the lens light, $I_{\rm L,limit} = 19.0$. We reject simulated events for which the lens brightness obey

$$M_I + 5\log\frac{D_{\rm L}}{10{\rm pc}} + A_{I,D_{\rm L}} < I_{\rm L,limit}, \quad (16)$$

where $A_{I,D_{\rm L}}$ is the extinction at $D_{\rm L}$, which we derive it following the procedure of Yang et al. (2022). We adopt the mass-luminosity relation of Wang et al. (2018),

$$M_I = 4.4 - 8.5\log\frac{M_{\rm L}}{M_\odot}. \quad (17)$$

The physical parameters derived from the Bayesian analysis are shown in Figure 6 and listed in Table 5, including the mass of the host star, $M_{\rm host}$, the planetary mass, $M_{\rm planet}$, the lens distance, $D_{\rm L}$, the projected planet-host separation, $r_\perp$, and the lens-source relative proper motion in the heliocentric frame, $\mu_{\rm hel,rel}$. We find that the preferred host star is a K or M dwarf located in the Galactic disk, with the disk lens probability of $P_{\rm disk} \sim 97\%$. The lens location is consistent with a direct estimate from a combination of the $\theta_{\rm E}$ distribution and the lens light constraint. With the $2\sigma$ lower limit of $\theta_{\rm E}$, 0.61 mas and a source distance of 8.5 kpc, we obtain $M_{\rm host} < 1.1\ M_\odot$ and $D_{\rm L} < 6.5$ kpc. The preferred planet is a Neptune-mass planet. The projected planet-host separation is $r_\perp = 1.9^{+0.6}_{-0.7}$ au for the "Close Central" solution and $r_\perp = 4.6^{+1.4}_{-1.7}$ au for the "Wide Central" solution. Assuming a snow-line radius $a_{\rm SL} = 2.7(M/M_\odot)$ au (Kennedy & Kenyon 2008), for the "Close Central" solution the planet is probably located near the snow-line, and for the "Wide Central" solution the planet is located well beyond the snow-line.

The estimated lens brightness of a main-sequence host is $I_{\rm L} = 20.4^{+1.4}_{-0.9}$ mag for both the "Close Central" solution and the "Wide Central" solution, which is similar to the source brightness. Bhattacharya et al. (2018) resolved the lens and the source of the event OGLE-2012-BLG-0950 when they were separated by about 34 mas using the Keck adaptive optics (AO) imaging and the *HST* imaging, for which the source and the lens also have roughly equal brightness. The estimated lens-source relative proper motion is $\sim 8$ mas yr$^{-1}$. Therefore, it may be possible to resolve the lens light using current instruments by about 2027 and can almost certainly be done using AO imaging on 30m-class telescopes, once they are available[3]. The high-resolution observations can also measure the lens-source relative proper motion $\mu_{\rm rel}$, which can yield the angular Einstein radius by $\theta_{\rm E} = \mu_{\rm hel,rel} \times t_{\rm E}$. The combination of the lens light and $\theta_{\rm E}$ can yield the lens mass and distance (e.g., Bhattacharya et al. 2018), which could be used to study the microlensing mass function, the microlensing planet frequency as a function of the Galactic environment, and the planet mass functions within different environments (Gould 2022).

---

[3] One exception would be if the host star were a stellar remnant (e.g., Blackman et al. 2021).

## 7 DISCUSSION

### 7.1 The Role of the Follow-up Data

In this paper, we have presented the analysis of KMT-2022-BLG-0440. Although the planetary signal is a short-lived and weak bump, the 2L1S "BD", 2L1S "resonant" and 1L2S solutions are excluded by the high-cadence, multi-band survey and follow-up data. To investigate the role of the follow-up data, we repeat the 2L1S and 1L2S analyses using only the survey data (KMTNet + MOA). We exclude the KMTNet and MOA data that were taken due to "auto-followup" for this high-magnification event.

Figure 7 shows a close-up of the anomaly with the survey-only data, for which the 2L1S parameters are listed in Table 6. The "central" solutions still provide the best fit for the observed data. The "resonant" and "BD" solutions are disfavored by $\Delta\chi^2 = 26.3$ and 17.6, respectively. If we adopt a significance level of $5\sigma$ as the rejection threshold, it is sufficient to break the "central-resonant" caustic degeneracy, while the "BD" solutions are strongly disfavored but not fully rejected. Moreover, as introduced in Section 2, many KMTNet images have elliptical PSF and on the *I*-band KMTNet BLG41 images the source is near a bleed trail. Thus, it would be somewhat questionable to exclude the "BD" and "resonant" solutions without the follow-up data.

For the 2L1S/1L2S degeneracy, we find that the 1L2S model without the "color argument" is favored by $\Delta\chi^2 = 8.0$ compared to the best-fit 2L1S solution, with $(V-I, I)_{\rm S2} = (2.31\pm0.34, 23.90\pm0.42)$. Then, the inclusion of the "color argument" only provides a $\chi^2$ increment of 1.6. Because the KMTNet "auto-followup" system also substituted the *V*-band observations, the KMTC41 and KMTS41 data together have eight *V*-band data points on the peak, which are twice the normal observations. Therefore, with the survey-only data the "color argument" is much weaker and would not permit us to rule out the 1L2S model.

The investigations above have two implications for future follow-up observations which are targeted at high-magnification events. First, simultaneous and "independent" high-cadence observations are necessary for confirming weak signals and breaking the degeneracy. LCO has eight 1.0 m telescopes at the same sites as KMTNet, so LCO + KMTNet can provide simultaneous follow-up observations. However, LCO and KMTNet should both take high-cadence observations if the observational resources are sufficient. For the present case, the severe photometric problem for the KMTC41 follow-up observations was discovered during the TLC reduction. Second, multi-band follow-up observations are important for breaking the 2L1S/1L2S degeneracy. For the present case, the $V$-band (from KMTNet) and $r$-band (from LCO) follow-up data play a decisive role in excluding the 1L2S model. The KMTNet "auto-followup" system may substitute more $V$-band observations for the follow-up target, and the LCO follow-up observations can be taken in the $r$ or even $V$ band if the target is bright and the extinction is not severe. As shown in Figure 2, the $r$-band LCO data are as accurate as the $I$-band LCO data at $I \sim 14.7$ mag.

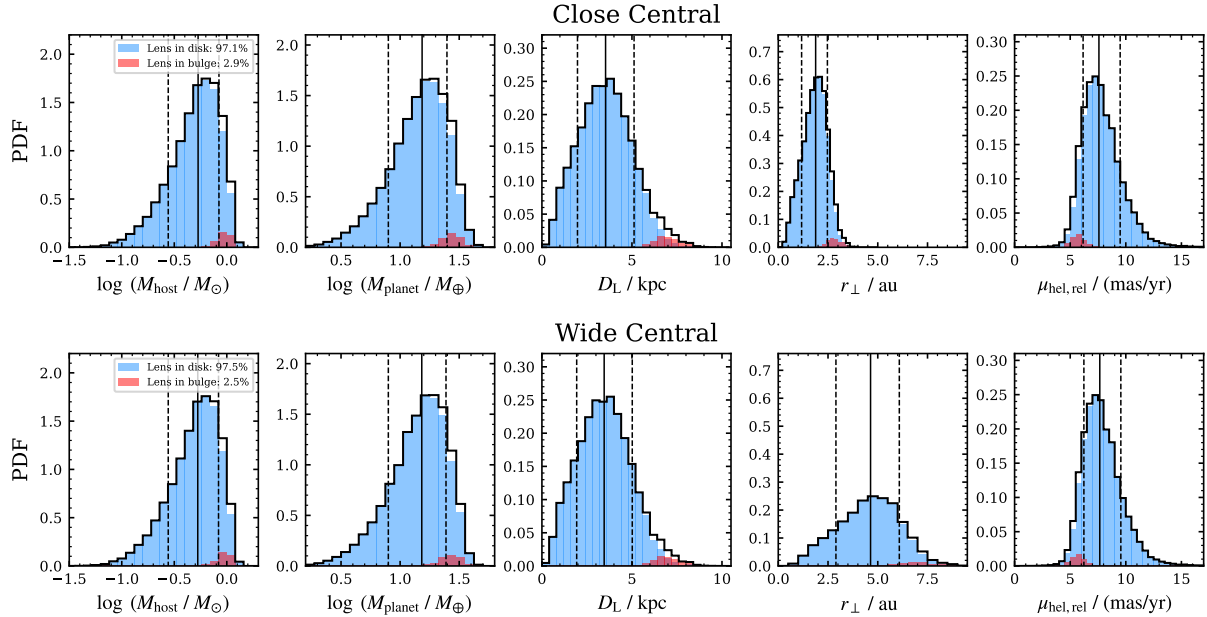

**Figure 6.** Bayesian posterior distributions of the mass of the host star, $M_{\rm host}$, the planetary mass, $M_{\rm planet}$, the lens distance, $D_{\rm L}$, the projected planet-host separation, $r_\perp$, and the lens-source relative proper motion in the heliocentric frame, $\mu_{\rm hel,rel}$. In each panel, the black solid line and the two black dashed lines represent the median value and the 15.9% and 84.1% percentages of the distribution, and the corresponding values are presented in Table 5. The distributions for the bulge and disk lenses are shown with red and blue, respectively.

### 7.2 The "Central-Resonant-BD" Degeneracy

Compared to the five 2021 events that suffer from the "central-resonant" caustic degeneracy (Yang et al. 2022; Ryu et al. 2022; Shin et al. 2023), the event KMT-2022-BLG-0440 has four differences. First, for the present case the "central-resonant" caustic degeneracy has been fully broken by the observed data with $\Delta\chi^2 = 91.1$. Second, the "BD" solutions provide better fits to the observed data than the "resonant" solutions. Third, for KMT-2022-BLG-0440 the $|\Delta\log(q)|$ between the "central-resonant" caustic degeneracy is 0.69 dex, while the five 2021 cases have $|\Delta\log(q)| \leq 0.28$. Fourth, for the present case the finite-source effects are marginally detected, while the five 2021 cases have unambiguous measurements on $\rho$ for all of the "central" and "resonant" solutions, with $\sigma(\rho)/\rho < 13\%$.

It is likely that the weak finite-source effects for the present case cause better fits for the "BD" solutions compared to the "resonant" solutions. For the present case and KMT-2021-BLG-1689 (Yang et al. 2022), which provided the parameters of the "BD" solutions, the finite-source effects of the "BD" solutions are marginally detected. As shown in Figure 4 of Yang et al. (2022), the BD solutions display a smoother anomaly, so if the finite-source effects are marginal, the BD solutions could be favored over the "resonant" solutions, which have caustic crossing for all of the six cases. We do not know whether the weak finite-source effects for the present case lead to the large $|\Delta\log(q)|$ and the large $\chi^2$ difference between the "central" and "resonant" solutions (even using the survey-only data). It would be worthwhile to conduct a literature search for the "Central-Resonant-BD" degeneracy and to re-analyze some published events. Such work may find missing degenerate solutions and thereby delineate more features for the "Central-Resonant-BD" degeneracy, which could improve theoretical understanding for the degeneracy.

For degenerate 2L1S solutions, Yang et al. (2022) proposed that the phase-space factors (Poleski et al. 2018) of $(\log s, \log q, \alpha)$ can be used to weight the probability of each solution. For the present case, the "resonant" and "BD" solutions have been fully excluded by $\chi^2$, but we still calculate the phase-space factors here. Following the procedures of Yang et al. (2022), we find that the "resonant" solutions are significantly disfavored by $p_{\rm resonant} : p_{\rm central} = 0.017$, similar to the two events reported by Yang et al. (2022). In addition, the "BD" solution is mildly favored by $p_{\rm BD} : p_{\rm central} = 2.8$.

### 7.3 The Third $q < 10^{-4}$ Planet from High-Magnification Planetary Signals

With a planet/host mass ratio, $q = 0.75$–$1.00 \times 10^{-4}$ at $1\sigma$, KMT-2022-BLG-0440Lb is a new $q < 10^{-4}$ microlensing planet. To investigate the roles in different periods of survey and follow-up observations in the detections of $q < 10^{-4}$ planets, we plot the $\log |u_{\rm anom}|$ versus $\log |u_0|$ distribution of all $q < 10^{-4}$ planets in Figure 8, where $u_{\rm anom}$ is the lens-source offset scaled to $\theta_{\rm E}$ at the time of planetary signals and $|u_{\rm anom}| \simeq |u_0/\sin(\alpha)|$. Red represents the planets detected by joint observations of surveys and follow-up, and black represents the planets solely discovered by surveys. Triangles and circles indicate the planets detected before 2016 and since 2016, respectively. Here we adopt the appearance date of a planetary signal as its discovery date, rather than the publication date.

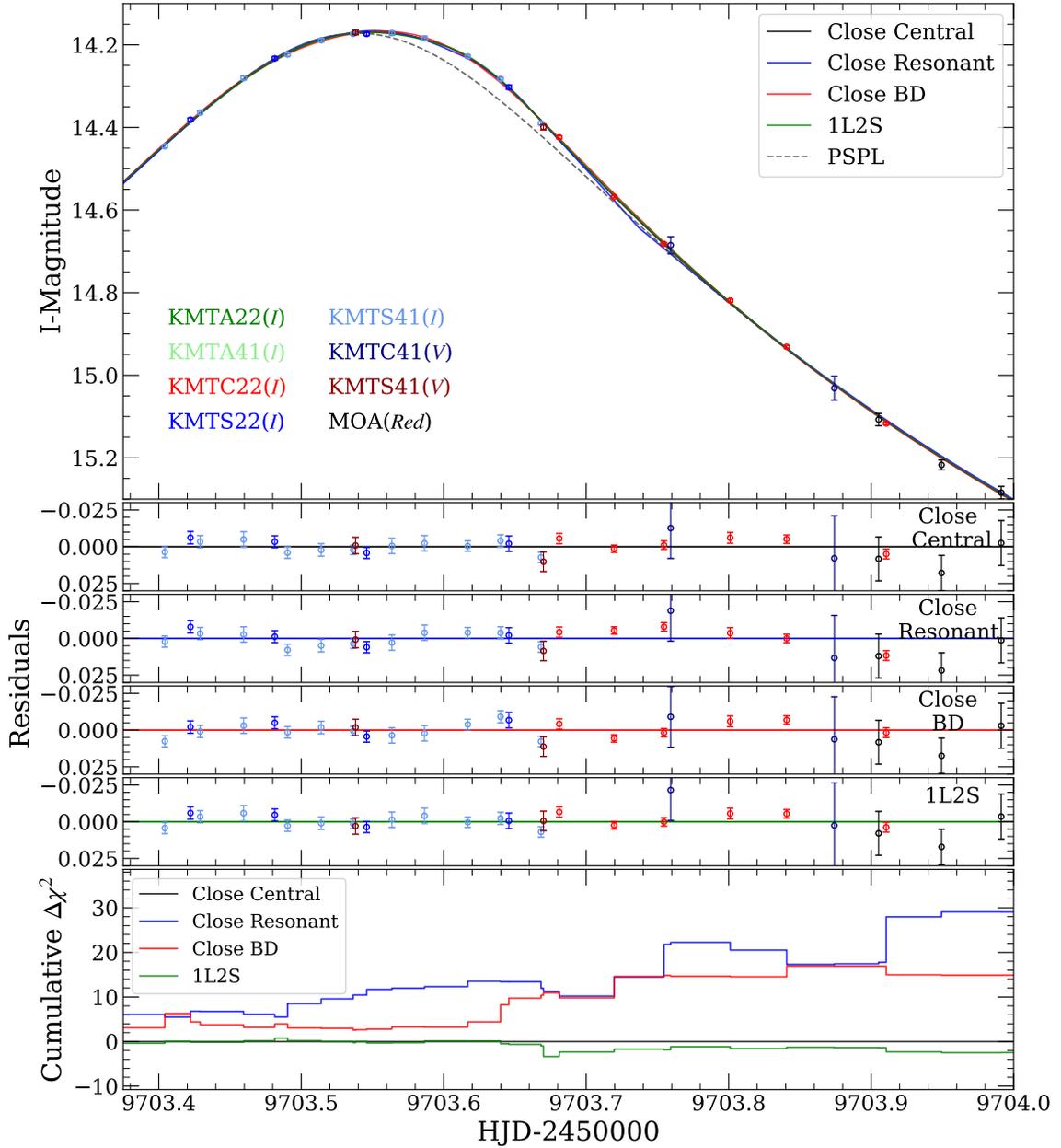

**Figure 7.** A close-up of the anomaly and models using only the survey data (KMTNet + MOA). The parameters of the 2L1S solutions are shown in Table 6.

Since the first $q < 10^{-4}$ planet, which occurred in 2005, only six $q < 10^{-4}$ planets were detected before 2016. Of them, the follow-up data played a major or decisive role in the first four planets from 2005 to 2009 (Beaulieu et al. 2006; Gould et al. 2006; Sumi et al. 2010; Muraki et al. 2011), and two were detected by survey-only data after the advent of the OGLE-IV survey in 2011 (Gould et al. 2014; Ranc et al. 2019). Since 2016, the advent of KMTNet has greatly enlarged the sample of microlensing $q < 10^{-4}$ planets. With the newly developed AnomalyFinder algorithm (Zang et al. 2021b, 2022), 27 planets with $q < 10^{-4}$ solutions were discovered from the first four-year (2016–2019) KMTNet survey (see Zang et al. 2023 and references therein). Another $q < 10^{-4}$ planet, OGLE-2018-BLG-0677Lb (Herrera-Martín et al. 2020), failed the detection threshold of the AnomalyFinder algorithm. Of the 28 planets, only two were detected by the survey + follow-up data. For the event OGLE-2019-BLG-0960, the follow-up data from the predecessor of the current follow-up program conducted by the MAP + KMTNet + $\mu$FUN teams, which observed the *Spitzer* microlensing events (Gould et al. 2018), played an important auxiliary role for this case (Yee et al. 2021). For the event OGLE-2018-BLG-1185, although the follow-up data covered the planetary signal, due to the dense data from the OGLE, MOA and KMTNet surveys the planet can be detected and characterized without the follow-up data (Kondo et al. 2021). That is, for about one decade (from 2009-mid to 2019-mid) the follow-up observations did not add any new $q < 10^{-4}$ planets besides the detections from the survey data, although during this period follow-up observations played a decisive role in many $q > 10^{-4}$ planets (e.g., OGLE-2015-BLG-0966, Street et al. 2016; MOA-2010-BLG-477, Bachelet et al. 2012). The possible reason is that the two follow-up teams, $\mu$Fun and the Probing Lensing Anomalies NET-

**Table 6.** 2L1S static models for KMT-2022-BLG-0440 using only the survey data

| Parameters | Close | | | Wide | | |
|---|---|---|---|---|---|---|
| | Central | Resonant | BD | Central | Resonant | BD |
| $\chi^2$/dof | 2262.5/2265 | 2296.7/2265 | 2280.1/2265 | 2262.7/2265 | 2288.8/2265 | 2280.1/2265 |
| $t_0 - 9703$ (HJD′) | 0.5396 ± 0.0007 | 0.5421 ± 0.0004 | 0.4249 ± 0.0156 | 0.5395 ± 0.0007 | 0.5420 ± 0.0006 | 0.4123 ± 0.0177 |
| $u_0 (10^{-3})$ | 4.0 ± 0.1 | 4.0 ± 0.1 | 1.5 ± 0.2 | 4.0 ± 0.1 | 4.0 ± 0.1 | 1.2 ± 0.3 |
| $t_E$ (days) | 41.80 ± 1.48 | 41.89 ± 1.50 | 41.31 ± 1.62 | 41.93 ± 1.46 | 41.63 ± 1.39 | 42.22 ± 1.47 |
| $\rho (10^{-3})$ | < 1.315 | 1.150 ± 0.222 | < 1.400 | < 1.230 | 1.214 ± 0.146 | < 1.227 |
| $\alpha$ (degree) | 300.12 ± 0.90 | 298.73 ± 1.02 | 42.54 ± 1.97 | 300.26 ± 0.95 | 298.84 ± 0.78 | 41.82 ± 1.71 |
| $s$ | 0.6048 ± 0.0379 | 0.9920 ± 0.0014 | 0.0990 ± 0.0103 | 1.6869 ± 0.1121 | 1.0126 ± 0.0026 | 11.3345 ± 1.2762 |
| $q (10^{-4})$ | 1.059 ± 0.307 | 0.197 ± 0.031 | 409.7 ± 88.3 | 1.111 ± 0.341 | 0.209 ± 0.025 | 493.9 ± 115.3 |
| $\log q$ | −3.992 ± 0.119 | −4.711 ± 0.070 | −1.398 ± 0.096 | −3.972 ± 0.121 | −4.684 ± 0.053 | −1.318 ± 0.101 |
| $I_{S,OGLE}$ | 20.186 ± 0.039 | 20.190 ± 0.040 | 20.170 ± 0.043 | 20.190 ± 0.038 | 20.183 ± 0.037 | 20.168 ± 0.039 |
| $V_{S,OGLE}$ | 22.003 ± 0.039 | 22.007 ± 0.040 | 21.988 ± 0.043 | 22.007 ± 0.038 | 21.999 ± 0.037 | 21.985 ± 0.039 |
| $I_{B,OGLE}$ | 18.232 ± 0.052 | 18.232 ± 0.052 | 18.235 ± 0.052 | 18.232 ± 0.052 | 18.233 ± 0.052 | 18.235 ± 0.052 |
| $V_{B,OGLE}$ | 19.996 ± 0.103 | 19.995 ± 0.103 | 19.998 ± 0.103 | 19.995 ± 0.103 | 19.996 ± 0.103 | 19.999 ± 0.103 |

NOTE. HJD′ = HJD − 2450000. The upper limit on $\rho$ is $3\sigma$. $t_0$ and $u_0$ take the position of the host star as the origin.

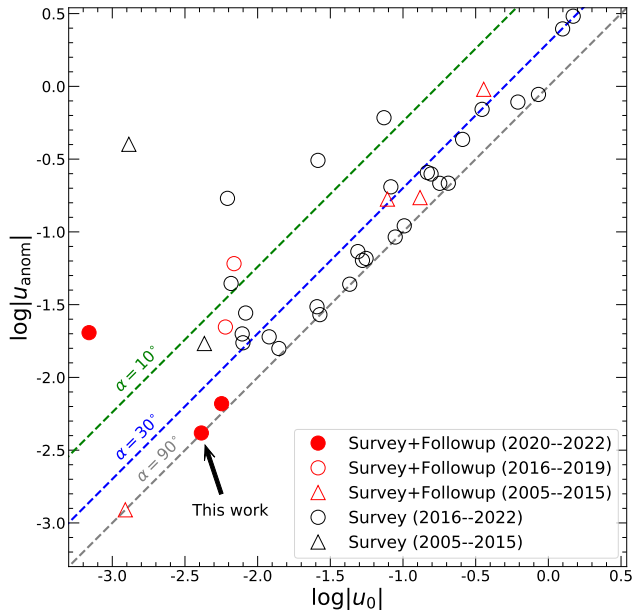

**Figure 8.** $\log |u_{anom}|$ vs. $\log |u_0|$ distribution for the 39 planetary events with $q < 10^{-4}$, adapted from Figure 15 of Zang et al. (2023). The black arrow indicates the planet discovered by this work. Red represents the planets detected by a joint observations of surveys and follow-up, and black shows the planets solely discovered by surveys. Triangles and circles indicate the planets detected before 2016 and since 2016, respectively. The green, blue, and grey dashed lines represent $\alpha = \arcsin(|u_0|/|u_{anom}|) = 10°$, $30°$, and $90°$, respectively, where $\alpha$ is the source trajectory with respect to the binary axis, and $u_{anom}$ is the lens-source offset in unites of $\theta_E$ at the time of planetary signals.

work (PLANET, Albrow et al. 1998), which were important in the detections of the first four $q < 10^{-4}$ planets, became less active over this decade.

Because most of the survey telescopes were shut down during the 2020 season and the AnomalyFinder algorithm has not been applied to the 2021 and 2022 KMTNet data, since 2020 only two survey-only $q < 10^{-4}$ planets have been discovered. Since 2020 July, the follow-up program conducted by the MAP + KMTNet + $\mu$FUN teams have found three $q < 10^{-4}$ planets. This detection rate is higher than the rate during 2005–2009 but is still significantly lower than the rate of KMTNet (5–8 per year, Zang et al. 2023). Of course, it is unfair to require the same productions from the KMTNet survey with ∼ 3000 hr/yr observing time of three identical 1.6 m telescopes and from a follow-up program with ∼ 700 hr/yr time of equally or less powerful telescopes. However, the productions of the current follow-up program were limited by four factors. First, the follow-up observations in 2020 were also affected by Covid-19. Second, the analysis of 2021–2022 follow-up data is incomplete. Third, the LCO observations conducted by the MAP collaboration had limited allocated time and a relatively low priority compared with other LCO observations. Fourth, about 1/3 of the KMTNet high-magnification events did not have follow-up observations due to the delay of the KMTNet AlertFinder system.

Figure 8 also shows that high-magnification events with $|u_0| < 0.01$ played an important role in the detections of $q < 10^{-4}$ planets, and 13 of the 39 planets were discovered from them. A notable feature of the $|u_0| < 0.01$ events is that, for these 13 events, the source trajectories are not uniformly distributed in the $\alpha$ space, with ten detected by oblique source trajectories with respect to the binary axis ($\alpha < 30°$) and five detected by very oblique source trajectories ($\alpha < 10°$). For planetary signals caused by central and resonant caustics, the oblique source trajectories have higher sensitivities to planets because the duration of a planetary signal is approximately $\propto (\sin \alpha)^{-1}$ and a planetary signal is stronger with a lower magnification for the underlying single-lens event (Yee et al. 2021). See Zang et al. (2021a) for an extreme case with $\alpha = 2°$.

Zang et al. (2021a) showed that for low-$q$ planets, dense or even continual observations are needed for perpendicular and nearly-perpendicular source trajectories. Zang et al. (2023) found a desert of high-magnification planetary signals ($A \gtrsim 65$) in KMTNet $q < 10^{-4}$ planetary sample. As shown in Figure 8, all of the five KMTNet $q < 10^{-4}$ planets with $|u_0| < 0.01$ have $\alpha < 30°$. Only three $q < 10^{-4}$ planets from high-magnification planetary signals ($A \gtrsim 65$) have been detected. The first $q < 10^{-4}$ planet, OGLE-2005-BLG-169Lb (Gould et al. 2006), was detected at $A \sim 800$, and 16 years later the follow-up program conducted by the MAP + KMT-Net + $\mu$FUN teams re-opened the window for high-magnification planetary signals for $q < 10^{-4}$ planets. Follow-up programs should keep dense or even continual observations for the high-magnification region because, intrinsically, more source trajectories can show planetary signals there despite the weak and short-lived planetary signals.

## ACKNOWLEDGEMENTS

We appreciate the anonymous referee for helping to improve the paper. J.Z., W.Zang, H.Y., S.M., S.D., Q.Q., Z.L., and W.Zhu acknowledge support by the National Natural Science Foundation of China (Grant No. 12133005). W.Zang acknowledges the support from the Harvard-Smithsonian Center for Astrophysics through the CfA Fellowship. This research has made use of the KMTNet system operated by the Korea Astronomy and Space Science Institute (KASI) and the data were obtained at three host sites of CTIO in Chile, SAAO in South Africa, and SSO in Australia. This research uses data obtained through the Telescope Access Program (TAP), which has been funded by the TAP member institutes. This research was supported by the Korea Astronomy and Space Science Institute under the R&D program (Project No. 2023-1-832-03) supervised by the Ministry of Science and ICT. This work makes use of observations from the Las Cumbres Observatory global telescope network. The MOA project is supported by JSPS KAKENHI Grant Number JSPS24253004, JSPS26247023, JSPS23340064, JSPS15H00781, JP16H06287, and JP17H02871. Work by C.H. was supported by the grants of National Research Foundation of Korea (2019R1A2C2085965 and 2020R1A4A2002885). Y.S. acknowledges support from BSF Grant No. 2020740. Work by J.C.Y. acknowledges support from N.S.F Grant No. AST-2108414. W.Zhu acknowledges the science research grants from the China Manned Space Project with No. CMS-CSST-2021-A11. The authors acknowledge the Tsinghua Astrophysics High-Performance Computing platform at Tsinghua University for providing computational and data storage resources that have contributed to the research results reported within this paper.

## DATA AVAILABILITY

Data used in the light-curve analysis will be provided along with publication.

This paper has been typeset from a T<sub>E</sub>X/L<sup>A</sup>T<sub>E</sub>X file prepared by the author.